\documentclass[pre,twocolumn,amsmath,amssymb,showpacs]{revtex4}
\usepackage{color}
\usepackage{graphicx}

\newcommand{\diff}{{\mathrm d}}

\begin{document}

\title{Ostwald Ripening in Multiple-Bubble Nuclei}

\author{Hiroshi Watanabe}
\email{hwatanabe@issp.u-tokyo.ac.jp}
\thanks{Corresponding author}
\affiliation{
The Institute for Solid State Physics, The University of Tokyo,
Kashiwanoha 5-1-5, Kashiwa, Chiba 277-8581, Japan
}

\author{Masaru Suzuki}
\affiliation{
Department of Applied Quantum Physics and Nuclear Engineering, Kyushu
University, 744 Motooka, Nishi-ku, Fukuoka 819-0395, Japan
}

\author{Hajime Inaoka}
\affiliation{
Advanced Institute for Computational Science, RIKEN,
7-1-26, Minatojima-minami-machi, Chuo-ku, Kobe, Hyogo, 650-0047, Japan
}

\author{Nobuyasu Ito}
\affiliation{
Department of Applied Physics, School of Engineering,
The University of Tokyo, Hongo, Bunkyo-ku, Tokyo 113-8656, Japan
}
\affiliation{
Advanced Institute for Computational Science, RIKEN,
7-1-26, Minatojima-minami-machi, Chuo-ku, Kobe, Hyogo, 650-0047, Japan
}

\date{\today}


\begin{abstract}
The ostwald ripening of bubbles is studied by molecular dynamics
simulations involving up to 679 million Lennard-Jones particles.
Many bubbles appear after depressurizing a system that is initially maintained in the pure-liquid phase,
and the coarsening of bubbles follows.
The self-similarity of the bubble-size distribution function predicted by Lifshitz-Slyozov-Wagner
theory is directly confirmed.
The total number of bubbles decreases asymptotically as $t^{-x}$ with scaling exponent $x$.
As the initial temperature increases, the exponent changes from $x=3/2$ to $1$,
which implies that the growth of bubbles changes from interface-limited (the $t^{1/2}$ law)
to diffusion-limited (the $t^{1/3}$ law) growth.
\end{abstract}

\maketitle

\section{Introduction}

When the pressure of a liquid is suddenly reduced, bubbles appear.
After the formation of bubbles, their coarsening takes place, \textit{i.e.}, larger bubbles grow at the expense
of smaller ones. This is known as Ostwald ripening. Ostwald ripening is one of the fundamental nonequilibrium phenomena and is commonly observed in many systems such as spin systems~\cite{Binder1977}, foams~\cite{Durian1991, Isert2013, Attia2013}, metallic alloys~\cite{Alloyeau2010, Werz2014}, and so forth. Therefore, the understanding of Ostwald ripening is of theoretical and practical importance. Lifshitz and Slyozov developed the theory of Ostwald ripening in the diffusion-limited case, which was followed by the theory of Wagner for the interface-limited case~\cite{Lifshitz196135,Wagner1961} (LSW). While the LSW theory has achieved great success qualitatively in predicting the behaviors of Ostwald ripening for various systems, it is highly nontrivial whether theory works for bubble systems.
There are two critical assumptions in the LSW theory.
One is the conservation law. The volume fractions of the second phase are assumed to be constant in the late stage of coarsening. When the second phase and the matrix are the same compounds,
the conservation law may not hold since the two phases can change between each other.
The other one is the mean-field treatment.
It has been reported in many studies that the cluster-size distribution is generally broader than that predicted by the LSW theory, for example, in alloy systems~\cite{Baldan2002}. This is because that the LSW treatment is a type of a mean field theory, which is not always justified~\cite{Voorhees1985, Baldan2002}. For the case of bubble growth in a liquid, this problem can be more severe since the interactions between bubbles via the ambient liquid are ballistic, and consequently, can be stronger than those between precipitates in alloy systems. Additionally, it is not clear that the growth processes of the bubbles are quasi-static, which is also assumed in the LSW theory. Therefore, the validity of the LSW theory should be verified for bubble systems. 
While a number of experiments on Ostwald ripening have been performed~\cite{Durian1991, Isert2013, Attia2013,Alloyeau2010, Werz2014,Tatchev2011}, few have focused on bubble nuclei, especially homogenous nuclei.
There have been studies, in which the nuclei of a small numbers of bubbles were numerically simulated by molecular dynamics (MD) simulations~\cite{Kraska2008, Matsumoto2008, Wang2009, Yamamoto2010, Chen2014}. However, there have been virtually no studies on the simulation of a multiple-bubble system with sufficient number of bubbles to study Ostwald ripening, mainly due to the lack of computational power.
Additionally, bubbles were initiated by hot spots in the past studies~\cite{Kraska2008, Wang2009}. Therefore, simulations of homogenous nuclei without initiation has been waiting.
Recently, it has become possible to perform a full-particle simulation of a multiple-bubble system by MD
owing to the development of computational power~\cite{fx10full}. 
With sufficient number of particles, which is typically more than a hundred million, one can obtain the bubble-size distribution with sufficient accuracy to make a detailed comparison between numerical results and the LSW theory. In the present paper, we perform MD simulations of the cavitation process in order to achieve the ideal homogeneous nuclei of bubbles and investigate the validity of the LSW theory for bubble nuclei.

\begin{figure}[bt]
\begin{center}
\includegraphics[width=8.5cm]{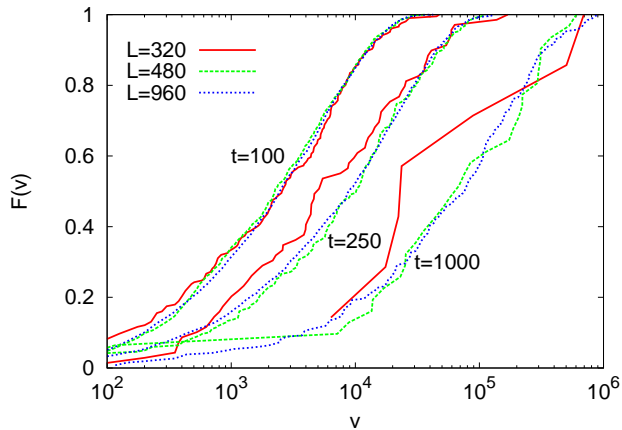}
\end{center}
\caption{
(Color online) The system-size dependence of CDFs defined in Eq.~(\ref{eq_cdf}).
Three system sizes are shown, $L=320, 480,$ and  $960$, respectively.
From left to right, CDFs at time $t = 100, 250, 1000$ are shown.
Decimal logarithms are taken for the horizontal axis.
}\label{fig_sdep}
\end{figure}

This paper is organized as follows. 
In Sec.~\ref{sec_lsw}, a brief overview of the LSW theory is given.
In Sec.~\ref{sec_method}, details of our method are described.
Numerical results are given in Sec.~\ref{sec_results}.
Finally, Sec.~\ref{sec_summary} is devoted to a summary and discussions of this study. 

\section{The Theory of Ostwald Ripening} \label{sec_lsw}

Consider a distribution function $f(v,t)$ that denotes a number of bubbles having volume $v$ at time $t$ in the system.
Several observables are defined as
\begin{eqnarray}
n(t) & \equiv& \int_0^{\infty} f(v,t) \diff v, \\
V_\mathrm{G}(t) &\equiv&  \int_0^{\infty} v f(v,t) \diff v, \\
\bar{v}(t) &\equiv& V_\mathrm{G}/n(t) ,
\end{eqnarray}
where $n(t)$, $V_\mathrm{G}(t)$, and $\bar{v}(t)$ are the total number of bubbles, the total volume of the gas phase, and the average volume of bubbles at time $t$, respectively.
The cumulative distribution function (CDF) is also defined as
\begin{equation}
F(v,t) \equiv n(t)^{-1} \int_0^v  f(v,t) \diff v. \label{eq_cdf}
\end{equation}
The time evolution of the distribution function is given by the following equation of continuity:
\begin{equation}
\frac{\partial f}{\partial t} = - \frac{\partial}{\partial v} \left(\dot{v} f \right), \label{eq_master}
\end{equation}
where $\dot{v}$ is a kinetic term that denotes volume changes of bubbles. The LSW theory assumes that the kinetic term depends only on the volume of bubbles $v$ at time $t$, \textit{i.e.}, all bubbles are subjected to identical pressure from the ambient liquid.
The critical assumption of the LSW theory is the self-similarity of the function forms as follows:
\begin{eqnarray}
f(v,t) &\sim& t^y \tilde{f}(v t^{-x}),  \label{eq_scale1}\\
\dot{v}(v,t) &\sim& t^{\omega} \tilde{\dot{v}}(v t^{-x}), \label{eq_scale2}
\end{eqnarray}
where $x, y$, and $\omega$ are scaling exponents.
It is also assumed that the total volume of the gas becomes almost constant in the long-time limit as
\begin{equation}
\frac{\diff V_G}{\diff t} = 0 \label{eq_constant}.
\end{equation} 
From the conservation law (\ref{eq_constant}) and the equation of continuity (\ref{eq_master}),
we obtain the scaling relations $y = -2x$ and $x = \omega +1$.
Then the asymptotic behaviors of observables are expected to be
\begin{eqnarray}
n(t) &\sim & t^{-x},\\
\bar{v}(t) &\sim & t^x, \\
F(v,t) & \sim & \tilde{F}(v t^{-x}).
\end{eqnarray}
This means that the asymptotic behavior is determined by only one scaling exponent $x$.
The scaling exponent $x$ is determined when the explicit expression of the kinetic term is given.

\section{Method}\label{sec_method}

To observe the time evolution of the distribution function $f(v,t)$, we perform MD simulations
with the truncated Lennard-Jones (LJ) potential of the form
\begin{equation}
V(r) =
\displaystyle 4 \varepsilon \left[
\left( \frac{\sigma}{r} \right)^{12} -
\left( \frac{\sigma}{r} \right)^{6} +
c_2 \left( \frac{r}{\sigma} \right)^2+ c_0 \right],
\end{equation}
with the well depth $\varepsilon$ and atomic diameter
$\sigma$~\cite{PhysRevA.8.1504}. The coefficients $c_2$ and $c_0$
are determined so that $V(r_c) = V'(r_c) = 0$ with the cut-off
length $r_c$ , i.e., the values of potential and force become
continuously zero at the truncation point and
$V(r)=0$ for $r > r_c.$
The truncation length is set to 3 throughout the simulations.
In the following, we use the physical quantities reduced by
$\sigma$, $\varepsilon$, and the Boltzmann constant
$k_\mathrm{B}$, e.g., the length is measured with
the unit of $\sigma$, and so forth.
We use kinetic temperature as temperature in the system, \textit{i.e.}, the temperature $T$
is defined by
\begin{equation}
T \equiv \frac{3K}{2},
\end{equation}
where $K$ is average kinetic energy per particle. The kinetic temperature can be defined
when the total energy of the system is conserved.

The system is a cube with the periodic boundary condition in all directions.
The time step is fixed to $0.005$ throughout the simulations.
We first maintain the system in the pure-liquid phase using the Nos\'e--Hoover thermostat~\cite{PhysRevA.31.1695}.
In a past study, the binodal line between the liquid and coexisting phases of this system was determined
to be $\rho_\mathrm{b}(T) =  a T + b + c (T_c - T)^{\beta}$, where 
$a = -0.195(1)$, $b = 0.533(1)$, $ c = 0.5347(4)$, $T_c = 1.100(5)$, and $\beta = 0.3285(7)$~\cite{ljuniv}.
Using these parameters, we set the initial density as $\rho = 1.04 \times \rho_\mathrm{b}(T)$, \textit{i.e.},
the initial density is set to 4\% higher than the coexisting density $\rho_\mathrm{b}$ at a given temperature.
The simulation conditions are listed in Table~\ref{tbl_conditions}.
After thermalization, the thermostat is turned off and the system becomes microcanonical.
The system is expanded as $L \rightarrow \alpha L$ and $\textbf{q}_i \rightarrow \alpha \textbf{q}_i$,
where $\alpha$ is a rescaling factor and $\textbf{q}_i$ denotes the position of particle $i$.
We chose $\alpha$ $ = 1.025$ for all runs.
After the expansion, the system is in the coexisting region in the phase diagram, and therefore,
bubbles appear as the result of spinodal decomposition. 
In order to study the time evolutions of bubbles, some criterion is necessary
to identify which particles are in gas or liquid state.
One possible definition is to take the Stillinger criterion~\cite{Stillinger1963};
two particles are in the same cluster when the distance between them are less than some threshold.
However, this criterion is too expensive for a system involving hundreds of millions of particles.
Instead, we use the subcell-dividing method in the present study, \textit{i.e.}, 
we divide the system into small subcells with length $3.075$, and
a subcell is defined to be in the gas state when its density is 
less than a density threshold.
The size of the subcells are chosen so that the subcell contains typically one particle
when the subcell is in the gas state. Based on our empirical basis, the size of subcells are small enough to resolve bubble configurations and large enough to identify phases between gas and liquid~\cite{PhysRevE.82.021201,fx10full}.
The densities of the gas and liquid coexisting in this system at the typical temperature $T=0.9$
are estimated to be $0.0402(2)$ and $0.6790(9)$, respectively~\cite{ljuniv}.
Here, we chose $0.2$ as the density threshold, \textit{i.e.}, a subcell having less
than six particles is defined to be in the gas state.
We confirmed that the results are insensitive to the density threshold~\cite{fx10full}.
The accuracy of bubbles' volume are determined by the volume of the subcells which is about 29 ($3.075^3$). In the scaling region, the average volume of the bubbles are larger than 5000. Therefore, the inaccuracy due to the subcell-dividing is at most 0.6\%. While this problem can be more serious when we consider the distribution function,
the inaccuracy due to the subcell-dividing is not serious as shown in Fig.~\ref{fig_cdf}.

We assume that the neighboring subcell in the gas state is in the same cluster
and identify the bubble on the basis of the site-percolation criterion in the simple cubic lattice.
We define clusters containing two or more subcells as bubbles.
The simulations are performed with a parallelized MD program~\cite{mdnote,fx10full,mdacp}.

We first investigate the system-size dependence. 
The time evolutions of CDFs of several system sizes are shown in Fig.~\ref{fig_sdep}.
Here, we consider three system sizes: $L = 320, 480,$ and $960$.
While three data are in good agreement with each other at $t = 100$, the data of the smallest size
becomes inaccurate at $t = 250$. The data of the middle size ($L = 480$) also becomes unreliable at $t = 1000$.
This means that we have to make the system as large as possible in order to make the scaling region of the LSW theory longer.
We confirmed that the distribution functions of the system with $L = 960$ show acceptable accuracy up to
$t = 10000$, \textit{i.e.}, the number of bubbles are enough to make sense to consider the distribution of them.
Therefore, we choose the size of the system to be $960$ throughout the simulations.

We use 4096 nodes of the K computer at RIKEN. We perform the simulation in the manner of the the flat-MPI.
Each run contains 32768 MPI processes.
After thermalization of $10^4$ steps, the observation is performed for $10^6$ steps.
The typical execution time of a single run is about 24 h.

\begin{table}
\begin{tabular}{|c|ccccc|}
\hline
$T$ & 0.8& 0.85 & 0.90 & 0.95 & 1.0 \\
$\rho$ & 0.767 & 0.735 & 0.7 & 0.66 & 0.613 \\
$N$ & 678592512 & 650280960 & 619315200 & 583925760& 542343168 \\
\hline
\end{tabular}
\caption{Initial conditions. The temperature $T$, density $\rho$, and number of particles $N$ are shown.
All systems are cubes with linear size $L=960$.
}\label{tbl_conditions}
\end{table}

\begin{figure*}[tb]
\begin{center}
\includegraphics[width=5.3cm]{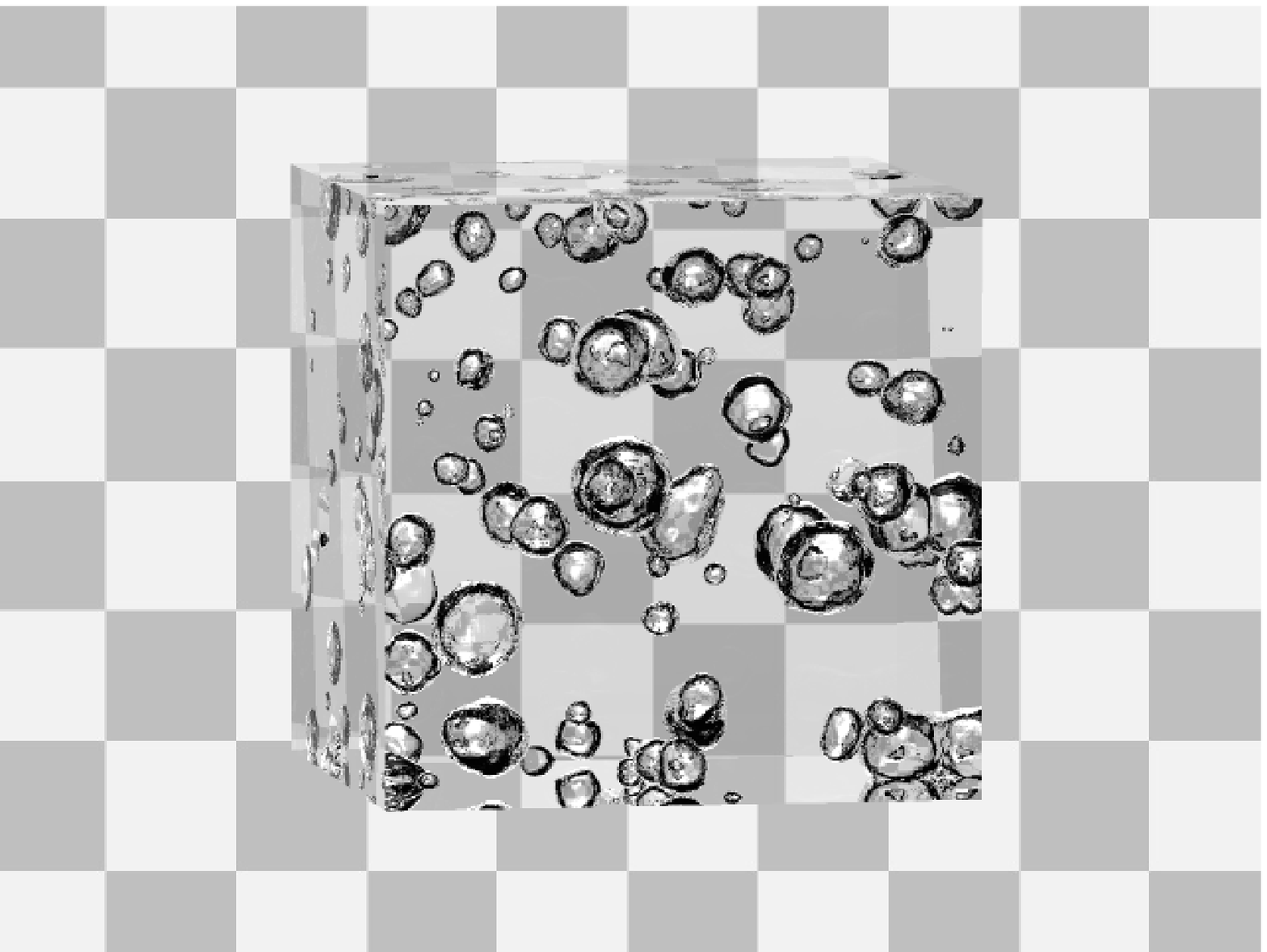}
\includegraphics[width=5.3cm]{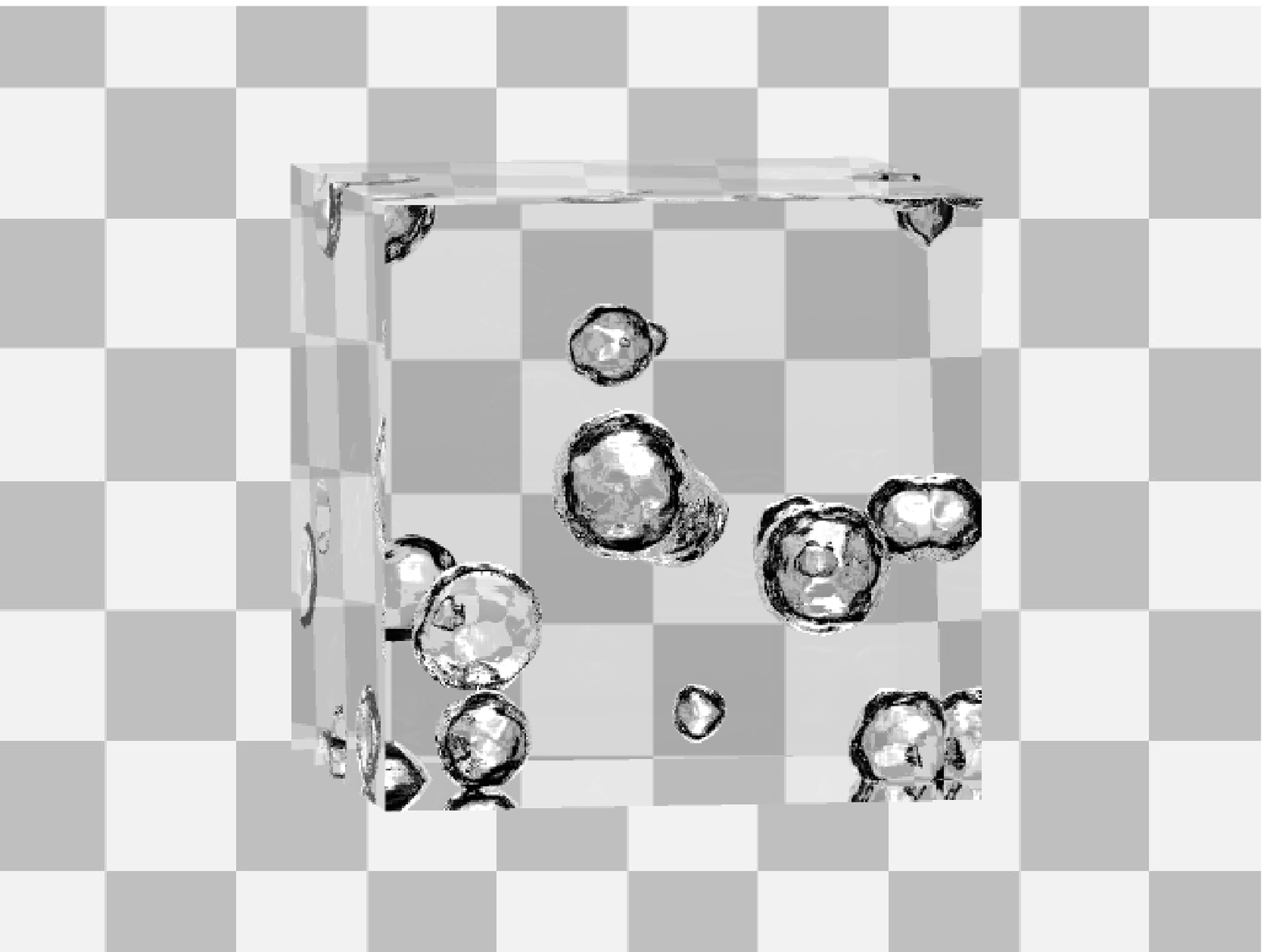}
\includegraphics[width=5.3cm]{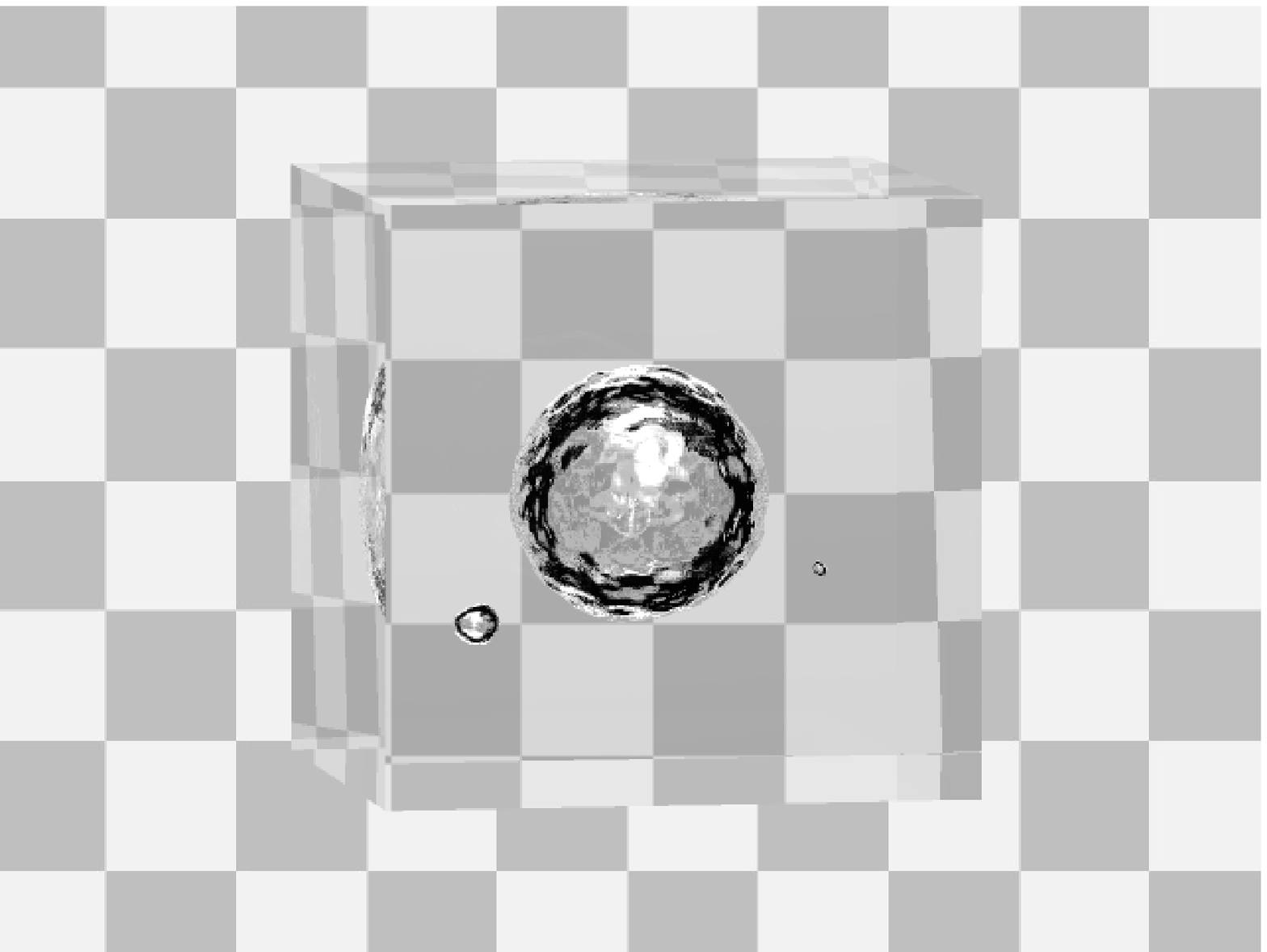}
\end{center}
\caption{
(Color online)  Time evolution of bubbles. A small system with $L=320$ is shown for visibility.
Left to right: snapshots at $t=50$, $150$, and $550$. The resolution of the snapshots is identical to subcells used in the analysis in the manuscript. Only the subcells are shown which are identified to be gas state.
}\label{fig_snapshots}
\end{figure*}

\begin{figure}[bt]
\begin{center}
\includegraphics[width=8.5cm]{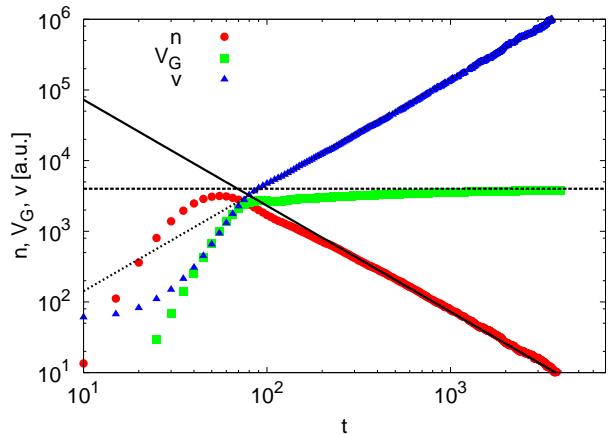}
\end{center}
\caption{
(Color online) Power-law behaviors of observables for $T=0.9$. The total number of bubbles $n(t)$, 
the total volume of gas $V_G(t)$, and the average volume of bubbles $\bar{v}(t)$ are shown, respectively.
The lines denotes, $t^{1.5}$, const., and $t^{-1.5}$.
Decimal logarithms are taken for both axes.
}\label{fig_power}
\end{figure}

\section{Results} \label{sec_results}
Typical snapshots are shown in Fig.~\ref{fig_snapshots} and
time evolutions of the observables at $T=0.9$ are shown in Fig.~\ref{fig_power}.
The total volume of gas $V_G(t)$ relaxes to its equilibrium value and becomes
almost constant for $t>100$ as assumed.
Therefore, the region $t>100$ can be regarded as the scaling region in the LSW theory.
Note that, the total volume of gas continues to increase in the scaling region while we assumed that
this is time-independent in Eq.~(\ref{eq_constant}). This increase in the total volume of the bubbles is due to the increase in temperature. Since the system in the scaling regime is non-equilibrium process, the temperature increases due to the entropy production. In the scaling regime, the total surface area decreases with almost keeping the total volume of bubbles. Then the total potential energy decreases and the total kinetic energy increases while the total energy (the sum of them) is always conserved after the expansion. Due to this effect, the temperature of the system slightly increases in the scaling regime. As temperature increases, the coexisting densities of gas and liquid changes. In the present system, the total gas volume slightly increases as temperature increases.

\begin{figure}[tb]
\begin{center}
\includegraphics[width=8.5cm]{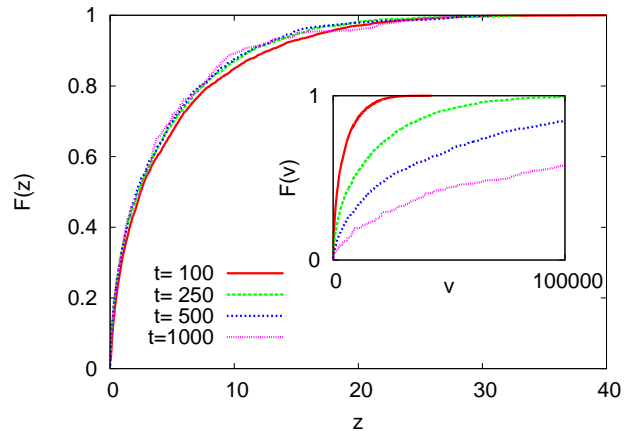}
\end{center}
\caption{
(Color online) Scaled cumulative distribution functions (CDFs) shown with respect to the scaling variable $z \equiv v t^{-x}$ where $x = 1.5$. (Inset) CDFs for $T=0.9$ at $t =100, 250, 500$, and $1000$.
}\label{fig_cdf}
\end{figure}

The volume fraction of the gas phase $V_G/L^3$, which is often denoted by $\phi$, in the scaling region is about $0.04$ for all cases.
While the number of bubbles $n(t)$ increases shortly after the expansion, it shows power-law decay
in the scaling region. The average volume of bubbles $\bar{v}(t)$ also shows power-law behavior in the scaling region.
The scaling exponent $x$ is determined to be $1.5$ which means that the average radius of the bubbles is proportional to $t^{1/2}$, \textit{i.e.}, the $t^{1/2}$  law is satisfied.
The CDFs and scaled CDFs are shown in Fig.~\ref{fig_cdf}.
They are well scaled using the exponent $x=1.5$. This is direct confirmation
that the distribution function has the asymptotic scaling form given by Eq.~(\ref{eq_scale1}).

\begin{figure}[htbp]
\begin{center}
\includegraphics[width=8.5cm]{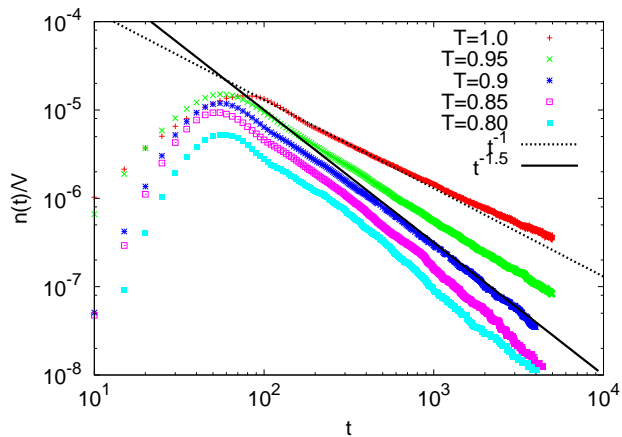}
\end{center}
\caption{
(Color online) Temperature dependence of the scaling exponent. The number densities of bubbles, $n(t)/V$, are shown
for $T=0.8, 0.85, 0.90, 0.95$, and $1.0$, where $V$ denotes the volume of the system.
The solid line and dashed line denote $t^{-1.5}$ and $t^{-1}$, respectively.
Decimal logarithms are taken for both axes.
}\label{fig_tdep}
\end{figure}

The temperature dependence of the scaling exponent is shown in Fig.~\ref{fig_tdep}.
As the temperature increases, a crossover from $x = 1.5$ to $1.0$ is observed.
When $x=1.0$, the average radius of the bubbles increases as $t^{1/3}$ (the $t^{1/3}$ law).
This crossover occurs at the threshold between interface-limited and diffusion-limited dynamics of the system.
While similar arguments have already been given~\cite{Lifshitz196135,Wagner1961, Binder1977}, we reformulate them in terms of bubble nuclei.
Consider a bubble with radius $R$ in an ambient liquid with pressure $p_\mathrm{l}$.
The inner pressure of the bubble is given by the Young-Laplace formula as
$p = p_\mathrm{l} + 2\gamma /R$ with surface tension $\gamma$.
The excess chemical potential of bubble with respect to the ambient liquid
is given by $\Delta \mu$.
If the evaporation/condensation rate is sufficiently high, then the difference in the chemical
potential is almost zero and the dynamics is governed by the diffusion process.
The diffusion current per unit area across the surface of the bubble
is given by Fick's law as
\begin{equation}
j = D \left. \frac{\diff \rho}{\diff r}\right|_{r=R} = \frac{D}{R}(C - \rho(R)),
\end{equation}
where $D$ is a diffusion constant and $C$ is a constant of integration.
The density at the surface of the bubble is given by the linearized Gibbs-Thomson equation as
$\rho(R) = \rho_\mathrm{eq} (1 - 2 \sigma \beta \gamma/ R )$,
where $\rho_\mathrm{eq}$ is the equilibrium density at the given temperature,
$\sigma$ is the atomic volume, and $\beta$ is the inverse temperature.
Since the growth rate is proportional to $4 \pi R^2 j$, we have
$\dot{v} \propto (v/v_c)^{1/3} - 1$, which leads to $\omega = 0$, and consequently, $x=1$,
\textit{i.e.}, the $t^{1/3}$ law is satisfied.
On the other hand, if the evaporation/condensation rate is much slower than the diffusion process,
then there is a finite gap in the chemical potential between the surface of the bubble and the ambient liquid.
We define a critical radius $R_c$ that makes $\Delta \mu = 0$.
Then the difference in the chemical potential for a bubble having radius $R$ is given by
\begin{eqnarray}
\Delta \mu & =& \int_{R_c}^{R} \frac{\diff \mu}{\diff p} \frac{\diff p}{\diff R} \diff R,\\
&=& - \int_{R_c}^{R} \frac{1}{\rho} \frac{2 \gamma}{R^2} \diff R,\\
&=& - \frac{1}{\beta}\int_{R_c}^{R} \frac{1}{(p_\mathrm{l} + 2 \gamma /R)} \frac{2 \gamma}{R^2} \diff R, \\
&=& \frac{1}{\beta} \ln \frac{1 + \delta/R}{ 1 + \delta / R_c}, \\
& \sim & \frac{\delta R_c}{\beta(R_c + \delta)} \left(\frac{1}{R_c} - \frac{1}{R} \right),
\end{eqnarray}
where $\delta \equiv 2\gamma / p_\mathrm{l}$.
Here, we used the Gibbs-Duhem equation $\rho \diff \mu = \diff p$ and the
ideal gas approximation $\rho = \beta p$.
Assuming that the growth rate of the bubble is proportional to $R^{d-1} \Delta \mu$ with the dimensionality of the system $d$, we obtain
\begin{eqnarray}
\dot{v} &\propto& R^{d-1} \left(\frac{1}{R_c} - \frac{1}{R} \right) ,\\
&=& v^{(d-2)/d} \left(\left(\frac{v}{v_c}\right)^{1/3} - 1 \right) ,\\
&\propto& t^{x(d-2)/d} \left(\left(\frac{v}{v_c}\right)^{1/3} - 1 \right). \label{eq_omega}
\end{eqnarray}
Comparing Eqs.~(\ref{eq_scale2}) and (\ref{eq_omega}), we have $\omega = x(d-2)/d$, and therefore, $x = d/2$.
For a three-dimensional system, $x$ is $3/2$,corresponding to the $t^{1/2}$ law.
The simulation results imply that the evaporation/condensation rate is much slower than
the diffusion process at low temperatures, and the reverse is true at high temperatures,
which makes intuitive sense.

\begin{figure}[tb]
\begin{center}
\includegraphics[width=8.5cm]{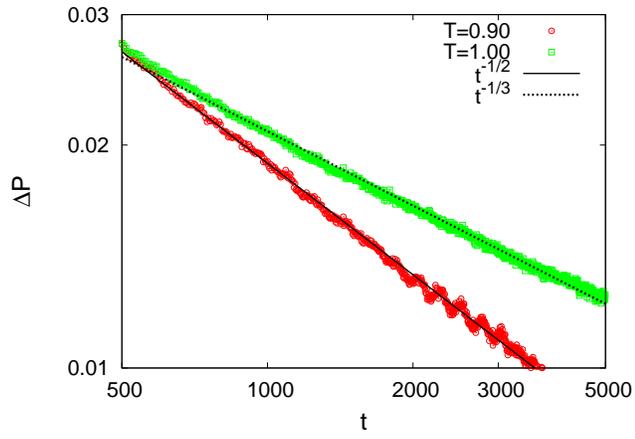}
\end{center}
\caption{
(Color online) Time evolution of pressure at $T=0.9$ and $1.0$.
The pressure difference $\Delta P \equiv P_0 - P(t)$ are shown.
The final pressures are estimated to be $P_0 = 0.02114(2)$ at $T=0.9$
and $P_0 = 0.05383(3)$ at $T=1.0$.
The solid line and dashed line denote $t^{-1/2}$ and $t^{-1/3}$, respectively.
Decimal logarithms are taken for both axes.
}\label{fig_pressure}
\end{figure}

So far, we only consider the physics at the surfaces of bubbles.
Making several approximations on the basis of the classical nucleation theory,
the asymptotic behavior of pressure can be derived~\cite{Binder1977}.
Suppose that $W(v)$ is the reversible work carried out to create a bubble having volume $v$. 
We assume that the function form of $W$ is that of the classical nucleation theory,
\begin{equation}
W(v) \propto \exp\left( \Delta h v  - \gamma v^{1-1/d}\right),
\end{equation}
where $\Delta h (t)$ is a time-dependent variable which is proportional to the chemical potential difference
between gas and liquid phases, $\gamma$ is a constant  which is proportional to the surface tension, respectively.
Neglecting the diffusive term, we have the Fokker-Planck-type equation (see, \textit{e.g.}, Eq.~3.39 in Ref.~\cite{Binder1977}):
\begin{equation}
\frac{\partial f}{\partial t} = - \frac{\partial}{\partial v }
\left\{
R f \left[ \Delta h - \gamma \left( 1 - \frac{1}{d} \right)v^{-1/d} \right]
\right\}, \label{eq_fp}
\end{equation}
where $R(v)$ is a quantity which depends only on $v$.
Considering the conservation law (\ref{eq_constant}), the scaling hypothesis (\ref{eq_scale1}), and
Eq.~(\ref{eq_fp}), we have the asymptotic behavior of $\Delta h (t)$ as
\begin{equation}
\Delta h(t) \sim t^{-x/d}.
\end{equation}
Then it is natural to expect that the asymptotic behavior of pressure also has the form
\begin{equation}
\Delta P(t) \equiv P_0 - P(t) \sim t^{-x/d},
\end{equation}
where $P(t)$ is the pressure of the system at time $t$ and $P_0$ is the final pressure,
\textit{i.e.}, $P_0 \equiv P(\infty)$, respectively.
The time evolutions of pressure at $T=0.9$ and $1.0$ are shown in Fig.~\ref{fig_pressure}.
We find that the asymptotic behavior of $\Delta P$ is $t^{-1/2}$ at the lower temperature 
and $t^{-1/3}$ at the higher temperature, which are consistent with the fact that
$x = 3/2$ at the lower temperature and $x =1$ at the higher temperature.

\section{Summary and Discussion} \label{sec_summary}

To summarize, we have performed MD simulations and observed the Ostwald ripening of bubbles.
To the best of our knowledge, this is the first study directly confirming the scaling behavior in multiple-bubble nuclei by MD simulations. At least a hundred million particles are required to perform scaling analyses of the distribution function with acceptable accuracy, which cannot be achieved without a peta-scale computer.

We have observed both interface-limited and diffusion-limited behaviors in the same system.
The scaling behavior predicted by the LSW theory is confirmed directly by observing the bubble-size distributions
and the asymptotic behaviors of pressure are also consistent with the theory.
It is rather surprising that the LSW theory works well for bubble nuclei, and this success is attributed to the separation of time scales. The LSW theory assumes that the coarsening rate of a bubble is independent of its surroundings.
This condition is justified by the separation of time scales between 
the relaxation time of the pressure in the ambient liquid
and the coarsening rates of bubbles.
The pressure of the system is almost homogeneous
throughout the time evolutions, and therefore, bubbles in the system are subjected to identical pressure throughout the time evolution, which justifies the mean field treatment.

The conservation of the total volume of the gas phase $V_\mathrm{G}$ also suggests the separation of time scales between time evolution of the volume and that of the surface area.
 As shown in Fig.~\ref{fig_power}, the total volume of gas increases quickly after the expansion,
and becomes almost constant. It implies that the time evolution of volume is much faster than that of surface area.
In the scaling region, the total surface area of the bubbles decreases keeping the total volume of bubbles constant.
As surface free-energy is released, the temperature of the system increases. This increase in temperature
is the driving force of Ostwald-ripening of bubbles. Note that, the total volume of gas increases slowly in the scaling region
due to the increase in temperature.

\section*{Acknowledgements}
The computation was carried out on the K computer at RIKEN.
We would like to thank N. Kawashima, H. Hayakawa, and S. Takagi for helpful discussions.
This work was partially supported by Grants-in-Aid for Scientific
Research (Contract No.\ 23740287).

\bibliography{ostwald}
\end{document}